# Tryptophan-containing proteins as label-free nanothermometers

Graham Spicer[1,2], Clara Maria Garcia-Abad[3,4], Alejo Efeyan[5] and Sebastian Thompson[3,4]*

___________

[1] Wellman Center for Photomedicine, Massachusetts General Hospital, 02114 Boston, MA, USA

[2] Harvard Medical School, Harvard University, 02115 Boston, MA, USA

[3] Madrid Institute for Advanced Studies in Nanoscience (IMDEA Nanociencia), C/Faraday 9, Madrid 28049, Spain

[4] Nanobiotechnology Unit Associated to the National Center for Biotechnology (CNB-CSIC-IMDEA), Madrid 28049, Spain

[5] Centro Nacional de Investigaciones Oncológicas (CNIO), 28029 Madrid Spain *Correspondence Sebastian Thompson, Fundación IMDEA nanociencias, 28049 Madrid Spain

Email: sebastian.thompson@imdea.org;

**Abstract**

There remains a need for techniques to monitor thermal processes at high spatiotemporal resolution, with myriad potential applications in chemistry, biology, and engineering. Measurement of temperature from nanoscale molecular phenomena are particularly promising due to their general compatibility with biological systems. Among these phenomena, fluorescence polarization anisotropy is particularly attractive due to its high sensitivity to thermal and photophysical information, and its general utility with a wide spectrum of fluorophores. In this work, we propose that the measurement of the intrinsic fluorescence polarization anisotropy of tryptophan can be used to measure label-free temperature at the nanoscale in a variety of tryptophan-containing proteins. We present a theoretical model of the temperature sensitivity of fluorescence polarization in free tryptophan and tryptophan-containing proteins, empirically explore these properties, and discuss the potential development of proteins as thermal memory sensors.

1. **Introduction**

With the aim of measuring temperature at the nanoscale, scientists have developed a vast number of nanothermometers in the last decade[1–6]. Nanothermometers can be used to sense temperature by monitoring the change in a measurable intrinsic physical or chemistry propriety. Among nanothermometers, particularly useful are optical nanothermometers because the interrogation of thermal environments can be accomplished non-invasively using light[7]. Prior embodiments of optical nanothermometers provided light emission through the addition or incorporation of a fluorescent dye as part of the nanothermometer structure[7]. This approach presents several disadvantages, including the challenge of introducing a fluorescent dye to the system to be studied without perturbing natural properties. To overcome this limitation, significant efforts to accomplish the measurement of temperature without the use of exogenous labeling[8] have been undertaken. Such measurements are based on the utilization of Raman spectroscopy to measure the temperature dependence of the O−H stretching Raman band in water, whereby the water molecule itself is utilized as an intrinsic nanothermometer. This methodology may be of utility for measuring local temperatures in a wide array of aqueous systems, but it also limited by the signal to noise of Raman spectroscopy, imposing a sensitivity limit to the temporal resolution of this approach when studying a dynamic system.

To overcome these limitations, we here report the first label-free protein nanothermometer based on the autofluorescence of tryptophan (Trp). The intrinsic fluorescent behavior of Trp allows the use of a wide array of Trp-containing proteins as nanothermometers by the measurement of the change in fluorescence polarization anisotropy with temperature. Anisotropy-based nanothermometers (ABNTs) present several advantages compared to other methods used for nanoscale temperature measurement. ABNTs can be used to extract temperature information from the fluorescence signal without any additional modification to the structure of the fluorescent protein or dye, thus enabling the observation of temperature information from all organelle trackers, dyes, fluorescent proteins (chimeras), coenzymes, and metabolism sensors/probes without affecting their innate activity and properties. Moreover, the ABNT is not sensitive to changes in absolute fluorescence intensity or quantum efficiency that may be caused by photobleaching, variations in illumination intensity, or fluorophore migration/concentration. Because the temperature sensitivity of the ABNT depends on the lifetime and the hydrodynamic behavior of the protein and the local environment of its fluorescent residues, this technique also can be applied to detect any changes in the 3D structure of the protein. Herein, we provide solid evidence for the universality of Trp containing proteins to be utilized as label-free nanothermometers.

## 2. Materials and Methods

### 2.1 Samples analyzed

The following proteins were purchased from Sigma Aldrich: α1-acid glycoprotein G9885, immunoglobulin G I5506, human albumin A3782, bovine serum albumin A7030, lysozyme L6876 and ribonuclease R6513, and dissolved in 1X PBS at a concentration of 1 mg/ml. DMEN (21063_029) was purchased from Gibco and FBS (SV30160.03) was purchased from Hyclone.

### 2.2 Temperature sensitivity measurements

For temperature sensitivity measurements calibration curve, fluorescence polarization anisotropy (FPA) versus temperature, was measured using a Horiba Fluorolog fluorometer running in T-format mode with vertically polarized excitation. Excitation and emission monochromators were set at 295 nm and 350 nm, respectively, to ensure only tryptophan was excited and not other autofluorescent residues[9]. Temperature was controlled using the temperature-control sample holder placed in the fluorometer. Each channel was corrected with horizontally polarized excitation. For each experiment, excitation and emission slits were selected to obtain around $10^6$ counts per seconds (cps). All samples were measured using a quartz cuvette.

## 3. Results

First, we studied the intrinsic fluorescence polarization anisotropy-based temperature sensitivity of Trp. Trp is an amino acid with a strong intrinsic fluorescence with excitation maximum at 295 nm and emission in the range of 310 nm to 380 nm. Free Trp in dilute aqueous solution was modeled with a theory previously introduced for rigid macromolecules[10,11] with several modifications required by the multiple fluorescence lifetimes of Trp[12]. The theoretical fluorescence polarization anisotropy (FPA) is expressed by the Perrin equation,

$$FPA = \frac{r_0}{1+\frac{\tau_f}{\Theta_r}} \quad (1)$$

where $r_0$ is the constant delimiting anisotropy equal to 0.4, $\tau_f$ is the fluorescence lifetime, and $\Theta_r$ is the rotational correlation time of the fluorophore. The effective Trp fluorescence lifetime, $\tau_f^{eff}$, is calculated from a weighted average of the two lifetimes reported for Trp,

$$\tau_f^{eff} = x_1\tau_{f1} + x_2\tau_{f2} \quad (2)$$

with values determined for Trp in aqueous solution reported in a previous work[13] indicating an explicit temperature dependence of $\tau_{f2}$ (13.6 °C = 3.75 ns, 25.6 °C = 2.3 ns, and 46 °C = 1.5 ns) but not $\tau_{f1}$ which has a fixed value of 0.5 ns. Trp's rotational correlation time can be expressed by the Stokes-Einstein-Debye relation,

$$\Theta_r = \frac{V\eta(T)}{kT} \qquad (3)$$

and depends on temperature, $T$, solvent viscosity $\eta(T)$ and its hydrodynamic volume $V$, or Stokes (hydrodynamic) radius which has been previously reported to be 0.34 nm (citation) ranging up to 0.50 nm depending on solution conditions. In the case of aqueous solutions, the solvent viscosity $\eta$ carries a temperature dependence given by the relation,

$$\eta(T) = \eta(20°C) \cdot 10^{\frac{20-T}{96+T} \cdot (A+B(20-T)+C(20-T)^2+D(20-T)^3)} \qquad (4)$$

where $\eta(20°C)$ is equal to 1.002 mPa·s, temperature $T$ is expressed in degrees Celsius, and the constants $A$, $B$, $C$, and $D$ have values of 1.2378, 1.303·10⁻³, 3.06·10⁻⁶, and 2.55·10⁻⁸, respectively[14]. From these relations the theoretical FPA is computed numerically in Matlab, with script available in Supplementary Information. Experimental measurements of Trp FPA in water were performed at temperatures of 20, 30, and 40 °C and are shown in Figure 1 for comparison to theoretical values from our model.

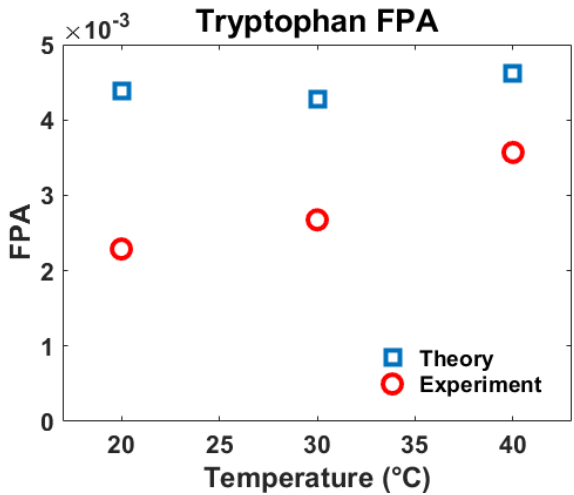

**Figure 1.** Theoretical and experimental values of FPA for Trp in solution.

To extend the implications of these results for temperature measurement, the differential derivative as a function of temperature provides the theoretical temperature sensitivity for Trp, which is shown in Figure 2.

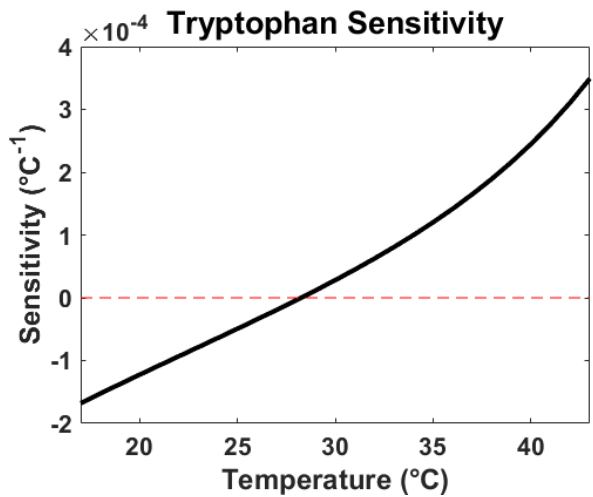

**Figure 2.** Theoretical thermal sensitivity of Trp in water.

This model demonstrates an exceedingly small thermal sensitivity of Trp FPA when dissociated in water over the physiologically relevant temperature range. Indeed, within this range, the predicted sensitivity is zero at 28 °C and remains below 0.2 mUnits of anisotropy per °C at 37 °C, providing poor temperature sensitivity, and corroborated empirically (not shown).

Nonetheless, when Trp is part of the amino acidic sequence of a protein, its intrinsic temperature sensitivity changes completely. For Trp fluorescence in proteins, the theoretical model is somewhat complicated by several salient features. First, Trp shows an additional fluorescence lifetime when compared to free Trp in solution. It is generally accepted that while the first two shorter fluorescence lifetimes of Trp are intrinsic to the structure of Trp, the larger third lifetime depends on the local environment and protein folding[15,16]. Second, proteins may contain multiple Trp residues, which each possess a unique local environment affecting the fluorescence lifetime and rotational degrees of freedom. These unique rotational degrees of freedom are not experimentally measurable for every protein.

To provide empirical support to the consequences of these effects, we conducted measurements of FPA as a function of temperature for several Trp-containing proteins with reported fluorescence lifetimes and hydrodynamic radii. The temperature sensitivity was calculated from the average change in FPA as a function of temperature from 20 to 40 °C. Table 1 shows the theoretical and experimental temperature sensitivity for several proteins, together with their lifetimes and hydrodynamic radii: human albumin (HA), bovine serum albumin (BSA), α1-acid glycoprotein (orosomucoid), lysozyme (Lyso), ribonuclease (Ribo), immunoglobulin G (IgG), and hydrolyzed casein (HyCase). Theoretical sensitivity values are calculated from the FPA given by equation (1) at the same measurement temperatures (20, 30, and 40 °C) while assuming a delimiting anisotropy $r_0$ of 0.4.

| Protein | $\tau_{f\,1}$ ns (Amplitude α) | $\tau_{f\,2}$ ns (Amplitud α) | $\tau_{f\,3}$ ns (Amplitud α) | Average $\tau_{f}$ ns | Rh (nm) | Theoretical Sensitivity | Practical Sensitivity |
|---|---|---|---|---|---|---|---|
| HA[16] | 6.55 (0.46) | 1.95 (0.23) | 0.17 (0.31) | 3.48 | 4.2 | 0.575 x$10^{-3}$ | 1.14x$10^{-3}$ |
| BSA[16] | 6.58(0.69) | 3.055(0.257) | 0.265(0.047) | 5.32 | 3.48 | 1.3 x$10^{-3}$ | 0.61x$10^{-3}$ |
| α1-acid glycoprotein[17] | 3.61 (0.28) | 1.42 (0.67) | 0.197 (0.07) | 1.97 | 2 | 1.8 x$10^{-3}$ | 0.31 x$10^{-3}$ |
| Ribo[16] | 3.32(0.77) | 1.18(0.23) | | 2.82 | 1.9 | 2.5 x$10^{-3}$ | 0.39 x$10^{-3}$ |
| IgG | NR | NR | NR | NR | 6.4 | -- | 0.2x$10^{-3}$ |
| Lyso[16] | 3.65(0.12) | 1.67(0.59) | 0.37(0.29) | 1.53 | 1.9 | 1.9 x$10^{-3}$ | 0.13x$10^{-3}$ |
| HyCase | NR | NR | NR | NR | <1 | 0 | 0.03 x$10^{-3}$ |

**Table 1.** Experimental and theoretic values of temperature sensitivity from selected proteins with known fluorescence lifetime and size.

Table 1 shows that the model of theoretical sensitivity does not match the experimental results, indicating a poor prediction of the general intrinsic temperature sensitivity from Trp residues in proteins with such simplified model. While additional parameters may be included to incorporate the relative proportion of segmental to bulk protein motions[18] or the effects of local Trp environment on spectral behavior of polarization excitation/emission and delimiting anisotropy[19], this task requires *a priori* knowledge that may not be available for any given protein, and thus limits its applicability. However, this potential shortcoming may be addressed through a general recommendation to perform experimental temperature sensitivity calibration for each protein of interest, as has been characterized for other proteins such as IgG in Table 1. Finally, when considering the measurements of hydrolyzed casein protein, the hydrolysate exhibited exceptionally low temperature sensitivity. This hydrolysate likely contains a polydisperse distribution of hydrodynamic radii, from large peptides (>7 amino acid residues) to single Tryp amino acids[20], along with a random local environment for the constituent Tryp residues in casein protein (1.25% w/w[21]). While this distributional behavior is not accounted for in the presented model of FPA temperature sensitivity, the extremely low sensitivity measured demonstrates that incorporation of the third fluorescence lifetime (influenced by protein folding) may be necessary to increase the thermal sensitivity. However, all the proteins show a good linear anisotropy dependence with temperature, a requirement to consider them as nanothermometers. Each curve is shown in Figure SI 1.

These theoretical and experimental results indicate the utility of a wide array of Trp-containing proteins as label-free nanothermometers. Using the intrinsic temperature sensitivity of BSA and HA as model proteins, we applied this phenomenon to two different applications. First, we explore the possibility of employing Serum proteins (the major one being BSA) as nanothermometer for cell culture experiments. Secondly, we demonstrate the possibility of using it HA as temperature memory sensor studying its thermal sensitivity after exposition to different temperature.

*Temperature sensitivity in FBS and complete cell culture medium.*
FBS includes all blood-borne proteins except for those involved in clotting and contained within blood cells. To assess the possibility of implementing FPA for temperature measurement in experiments where cell culture media with similarly complex molecular environments are used, we studied the temperature sensitivity of the complete cell culture medium with FBS at a concentration of 10% w/w. The temperature sensitivity of both FBS and complete cell culture medium is conferred by the Trp containing protein components of FBS, the major one being BSA[22].

The temperature sensitivity of FBS, complete cell culture medium (with FBS), and cell culture medium (no FBS) is illustrated in Table 2:

|  | Sensitivity |
|---|---|
| FBS | $0.41 \times 10^{-3}$ |
| Medium + FBS | $0.85 \times 10^{-3}$ |
| Medium | $-0.03 \times 10^{-3}$ |

**Table 2.** Intrinsic temperature sensitivity of FBS and cell culture medium.

Both FBS and complete medium present appreciable temperature sensitivity. This sensitivity is readily detectable with current polarization-sensitive fluorimeters and microscopes, opening the possibility to measure temperature in blood or any other fluid in living cells and organisms. As a negative control, culture medium alone, lacking the proteins in FBS, shows no thermal sensitivity.

*Temperature and protein structure as a memory sensor*
An exciting application for this temperature sensitivity is the possibility to monitor changes in the structure of proteins. To demonstrate this application, we measured the FPA temperature sensitivity of BSA before and after exposure to heat. BSA undergoes denaturation and aggregation (changing its hydrodynamic radius) when exposed to temperatures between 65 °C and 80 °C[23]. Figure 3 shows the experimental design to probe BSA temperature sensitivity as a function of heat exposure, along with resultant FPA sensitivity measured.

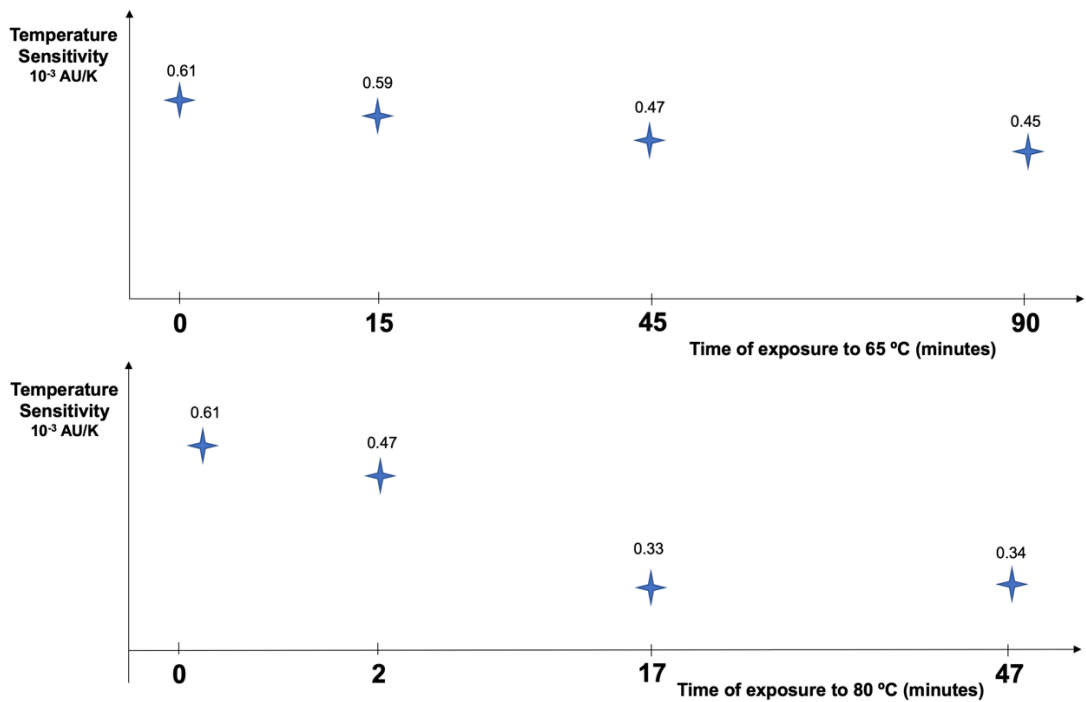

**Figure 3.** Experimental design for characterization of the response of FPA temperature sensitivity to heat-mediated denaturation of BSA.

Table 3 illustrates the degree of denaturation, the hydrodynamic radius of the protein, and the measured temperature sensitivity. It should be noted that the temperature sensitivity is measured after the protein has been exposed to the elevated temperature for the given time.

| Temperature/time | Degree of denaturation [citation] | Hydrodynamic Radius (nm) | Temperature Sensitivity (Anisotropy Units/K) |
|---|---|---|---|
| No exposure | 0 | 3.48 | $0.61 \times 10^{-3}$ |
| 65 °C for 15 min | 0.58 | 6 | $0.59 \times 10^{-3}$ |
| 65 °C for 30 min | 0.8 | 8 | $0.47 \times 10^{-3}$ |
| 65 °C for 60 min | 1 | <10 | $0.45 \times 10^{-3}$ |
| 80 °C for 2 min | 0.45 | 5 | $0.47 \times 10^{-3}$ |
| 80 °C for 15 min | 0.9 | 18 | $0.33 \times 10^{-3}$ |
| 80 °C for 30 min | 1 | <20 | $0.34 \times 10^{-3}$ |

**Table 3.** Temperature sensitivity indicates degree of protein denaturation.

The strong anticorrelation of FPA temperature sensitivity and degree of protein denaturation indicates that this optical measurement may be applicable for detecting any change in protein structure (as a non-invasive, complementary technique to traditional chemical and binding-based assays used to measure degree of protein denaturation). While both irreversible aggregation and denaturation occur during heating, temperature sensitivity is a strong indicator of prior exposure to elevated temperature. This opens the possibility for combination of FPA measurement with a sentinel protein of interest to indicate prior exposure to past temperature elevation in biological samples or shipping materials.

To further confirm that FPA temperature sensitivity can detect any change in protein structure in the absence of aggregation, we exposed HA to a chemical denaturant, 6M Guanidine Hydrochloride (GH). The results of this treatment are illustrated in Table 4.

| Protein | No denaturation | Denatured (6M GH treatment) |
| --- | --- | --- |
| HA | 0.01097 | 0.003098 |

**Table 4.** FPA temperature sensitivity of HA before and after chemical denaturation.

## 4. Discussion

Herein we demonstrate that any Trp-containing protein can be used as a label-free nanothermometer. Furthermore, we present evidence that FPA-based temperature sensitivity measurements of Trp-protein nanothermometers are sensitive to changes in protein structure, such as denaturation. While this information may confer limitations to the study of certain biological systems that undergo structural modifications that are not well-characterized, this system also has the potential to couple nanothermometry measurement with a sentinel protein to detect past exposure to elevated temperature. This protein-based thermal sensing approach can detect any changes in the structure of the protein due to thermal or chemical denaturation.